\documentclass[12pt,a4paper]{article}
\usepackage{lipsum}
\usepackage{authblk}
\usepackage[top=2cm, bottom=2cm, left=2cm, right=2cm]{geometry}
\usepackage{fancyhdr}
\usepackage{graphicx}
\usepackage{amsmath}
\usepackage{amsfonts}
\usepackage{amssymb}
\usepackage{subfig}
\renewenvironment{abstract}{%
  \hfill\begin{minipage}{0.95\textwidth}
  \rule{\textwidth}{1pt}}
{\par\noindent\rule{\textwidth}{1pt}\end{minipage}}
\makeatletter
\renewcommand\@maketitle{%
  \hfill
  \begin{minipage}{0.95\textwidth}
  \vskip 2em
  \let\footnote\thanks 
  {\LARGE \@title \par }
  \vskip 1.5em
  {\large \@author \par}
  \end{minipage}
  \vskip 1em \par
}
\makeatother
\begin{document}
%
%title and author details
\title{\textbf{Dynamic Regulation of T cell Activation by Coupled Feedforward Loops}}
\author[1,2]{Gershom Buri}
\author[3]{Girma Mesfin Zelleke}
\author[1]{Wilfred Ndifon}
\affil[1]{African Institute for Mathematical Sciences, Cape Town, 
  South Africa.}
\affil[2]{SACEMA, Stellenbosch University, Cape Town, South Africa.}
\affil[3]{African Institute for Mathematical Sciences, Limbe,
  Cameroon.}
\maketitle
\begin{abstract}
\pdfoutput=1
The adaptive immune system responds to foreign invaders by activating T cells. However, this
response is perilous if T cells are activated by the wrong signal or if they remain activated
unduly long after the threat has been eliminated. It is therefore important that T cells get activated only by the right kind of signals and
for the right duration. The dominant theory in immunology over recent decades has been that
a T cell must receive at least two signals before it can become activated. It is, however, unclear
whether and how this two-signal requirement ensures that T cell activation is provoked only by
the right signals and that  the response is just long enough. Here, we propose that the two
signal requirement induces a novel coherent feedforward motif whose properties align with those
preferred for an ideal immune response. Further consideration of the interaction between helper 
and regulatory  T cells induces a composite feedforward-feedback motif, analysis of which generates a condition   for a healthy  concentration of activated T helper cells.
\end{abstract}
\section{Introduction}

The immune system serves  to defend the body
against attacks by both foreign and internal threats. 
At the heart of an adaptive immune response lies the activation of T cells, the main orchestrators of the immune response. If not well regulated, this process  can have perilous consequences. 
Both a prolonged response and a response to  a  wrong target can   lead to destruction of  body tissue and have been  linked to pregnancy
complications, stroke, heart attacks, and  blood clots \cite{davidson2001autoimmune}.
In addition, they can lead to wastage of resources such as energy used in the production of cytokines \cite{rauw2012immune, lee2006linking}. 
What then determines whether a particular T cell is activated by the right signal, and that the activation doesn't last too long?

T cell activation is triggered by the binding of a T cell receptor (TCR) to its cognate antigen.  
However, in the absence of costimulation, TCR mediated recognition   
has been reported to cause a state of nonresponsiveness called anergy \cite{schwartz1992costimulation}. The two-signal theory posits that  T cell activation requires at least two signals: one signal in the form of antigen recognition by the TCR and the other in form of a secondary stimulus. 
This  requirement  was first proposed  as a mechanism for distinguishing the self from the non-self  in B cells \cite{bretscher1970theory}. The theory was later extended to T cell activation  by Lafferty and Cunningham  \cite{lafferty1975new}, and was reinforced  by the identification of the CD28 receptor together with the B7-1 molecule as its ligand \cite{june1987t,linsley1990t}.
Besides CD28, additional costimulatory molecules such as  ICOS, OX40 and 4-1BB%, and with  their ligands ICOS-L, OX40L and 4-1BBL
, have since been identified \cite{so2008immune,hutloff1999icos}. The complexity of  T cell activation, however, became more apparent with the discovery of coinhibitory receptors such as CTLA-4 and PD-1. These receptors  are expressed on the surfaces of activated T cells and down-regulate  T cell activation.  
The absence of these coinhibitory signals in mice has been associated with autoimmunity,  highlighting their possible role in maintenance of self tolerance \cite{waterhouse1995lymphoproliferative,nishimura2001autoimmune}. Despite   this advance in knowledge about costimulation, the entirety of the costimulation spectrum is still  yet to be  uncovered. 

Amongst the key cellular players of the immune system, with antagonistic roles, are helper and regulatory T cells. Helper T cells serve to coordinate the effector function of other white blood cells, while regulatory T cells down-regulate immune responses induced by helper T cells. In  so doing, regulatory T cells play a critical role in preventing over-reaction, and in  maintaining peripheral tolerance by suppressing activation of self reacting T helper cells \cite{sakaguchi2008regulatory, tang2008foxp3}. The two cell types are activated alike: 
Costimulation was initially defined solely with respect to helper T cells \cite{mueller1989clonal, lenschow1996cd28}. 
However, shortly after
the renewal  of interest in regulatory T cells (previously dismissed), the revelation of the necessity of costimulation in their
activation played a critical role in our understanding of the mechanisms of costimulation \cite{salomon2000b7,sansom2006role,tang2003cutting}.
This implied that some costimulatory molecules,
for example CD28 and IL-2, had dual (opposing) effects on immune responses, 
which made the prediction of therapeutic outcomes aimed at blocking any of these signaling pathways difficult \cite{lenschow1996cd28, bouguermouh2009cd28,adams2016costimulation}.   
This also confounded the determinants of the  relative amount of activated regulatory vs helper T cells-- a significant determinant of the outcome of an immune response \cite{bour2004costimulation,rossini1999induction, salomon2000b7}

Here, we propose to study T cell activation from the perspective of  network science. Despite the discovery
of a number of other cosignalling molecules, signal transduction between CD28 receptors and the B7
molecules remains the best defined initiator of T cell activation \cite{chen2013molecular, smith2009t}. We initially restrict our attention to this classical version of the two-signal theory with plans of
considering the dynamic interplay between costimulation and coinhibition
in the future. 
We show that the two-signal requirement induces a novel type-1 coherent feedforward loop. Feedfoward loops  are amongst the most significant motifs found in transcription networks of  diverse  gene systems \cite{mangan2003coherent,milo2002network,lee2002transcriptional}, and in other biological networks \cite{white1985neuronal, alberts2013essential}.     Next, we analyze the implications of this model of T cell activation on the interaction between  helper and regulatory T
cells. This results in a coupled coherent-incoherent feedforward  interaction loop. Sontag \cite{sontag2017dynamic} recently mathematically analyzed a similar interaction   by combining  negative feedback and  type 1 incoherent feed forward loops  to model immune responses to antigen presentation. In addition to helper and regulatory T cell nodes, his model includes a pathogen node that we don't consider here. Sontag \cite{sontag2017dynamic} also considers  dynamic rather than static antigen presentation as we have done here. As a result of these differences in model formulations, the two approaches to modeling immune responses  yield different contextual outcomes.

\section{Materials and Methods}
Briefly, a typical immune response  begins with the  engulfing, and then processing, of  infectious pathogens by  antigen presenting cells (APCs) of the innate immune system: mainly dendritic cells and macrophages. Peptide fragments of the pathogens are then bound to MHC class II molecules and  displayed for   CD4 T cell recognition \cite{nakagawa1999impaired}.  Upon successful recognition of such peptide-MHC complexes (pMHCs), the T cells are activated, and they go on to orchestrate the rest of the immune response against the infection. 
Once the infection has been  nullified, suppressor T cells known as T regulatory cells turn off
the immune response \cite{abbas2000immunology}. In this work, we studied the dynamics of T cell activation, and the interaction between  helper and regulatory T cells,  using mathematical models inspired by recent advances in network sciences.
For convenience, we  collectively refer to  CD4 and CD8 T cells as effector T cells.

\subsection{T cell activation induces a type-1 Coherent Feedfoward Loop (C1-FFL) with AND logic}
According to the two signal theory, one signal for T cell activation is in the form of TCR-pMHC binding whilst the other is mediated by costimulatory proteins on the surfaces of T cells and APCs.
The B7 molecules are only expressed on professional APCs, and their production peaks after the APCs
have been activated by microbial products \cite{abbas2000immunology}. The combined action of the two signals
activates T cells to proliferate and begin to differentiate into effector T cells \cite{abbas2000immunology, nijkamp2006principles}.

\begin{figure}
\center
\includegraphics[scale=0.4]{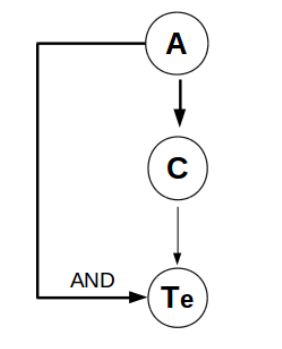}
\caption{Structure of the coherent type 1 feedforward loop proposed to govern T effector cell ($Te$) activation. The presence of an antigen ($A$) upregulates production of costimulation signals ($C$). Both $A$ AND $C$ induce activation of $T_e$.}
\label{fig:b1}
\end{figure}

From the description above, T cell activation induces a feedfoward loop (FFL) in which the antigen signal ($A$) regulates the costimulatory signal ($C$), and both jointly regulate T cell ($T_e$) activation (Figure \ref{fig:b1}). In this particular case, the FFL is ``coherent"  because the sign of
the direct regulation path (from $A$ to $T_e$) is same as the overall sign of the indirect 
path  through $C$. This FFL  motif is of type-1 because all the paths are positive (activators). Also, since both signals are required to activate the T cell, the input function of $T_e$ activation follows an AND logic gate. The other alternative is an OR-gate which obtains when either of the two signals is sufficient to activate the T cell.
\subsubsection{Mathematical model of the type-1 coherent feedforward loop (C1-FFL)}
\label{sec:sec211}
Using a binary (ON-OFF) antigenic signal $A$,, the rates of 
change in the production of  $C$ and $T_e$ were described by 
\begin{align}
\label{eqn:eqn1}
\begin{split}
\frac{dC}{dt} &= \beta_c \frac{A^n}{K^n_{AC} + A^n} - \alpha_cC,\\
\frac{dT_e}{dt} &= \beta_{T_e}\bigg( \frac{A^n}{K^n_{AT_e} + A^n} \times \frac{C^n}{K^n_{CT_e} + C^n}\bigg) - \alpha_{T_e} T_e,
\end{split}
\end{align}
where the parameters $K_{AC}$, $K_{AT_e}$ and $K_{CT_e}$ are the activation coefficients of $C$ by $A$, $T_e$ by $A$, and $T_e$ by $C$, respectively, while $\beta_c$ and $\beta_{T_e}$ are the maximal production rates  of the costimulatory signal and the T cell, respectively. We assumed that the production 
of $C$ and $T_e$ are offset by the decay rate parameters $\alpha_c$ and $\alpha_{T_e}$, respectively.

The general form,
\begin{eqnarray}
\label{eqn:23}
f(X) = \frac{X^n}{K_{XY}^n+X^n}
\end{eqnarray}
in Equation \ref{eqn:eqn1}
is called the Hill function and it represents the probability of activation of  $Y$ for  a given  concentration of $X$. The parameters $K_{XY}$ and $n$ represent the activation coefficient and the Hill coefficient, respectively. The activation coefficient represents the ligand concentration that generates a 50$\%$ chance of activation, while the  Hill coefficient represents the effective number of molecules of $X$ required to activate $Y$. When $X>>K_{XY}$, $f(X)\rightarrow 1$ and when $X<<K_{XY},$ $f(X)\rightarrow 0$. The higher the  value of $n$, the more the Hill function behaves like a step function (Supplementary Figure 1). %\ref{fig:b7}). 
In this scenario, $K_{XY}$ behaves like a threshold value for the ON and OFF regulatory influence of $X$. 	When $n$ is close to zero however, the Hill function can still be approximated  by the logic function
\begin{eqnarray}
f(X) = \theta(X>K_{XY}),
\end{eqnarray} 
which is equal to one when the concentration is greater than a particular activation threshold $K_{XY}$, and zero otherwise.
The equivalent form of Equation \eqref{eqn:23} for the inhibition of $Y$ by $X$ is given by
\begin{eqnarray}
\label{eqn:22}
f(X) = \frac{K_{XY}^n}{K_{XY}^n+X^n}.
\end{eqnarray}

Consider a case when the antigenic signal is ON for a time, $t_1\leq t< t_2$. If this is the initial instance of the infection, then until $t=t_1$, the costimulation signal $C$ is off, i.e, is zero. At $t=t_1$, $A$ switches ON (A = 1) and the dynamics of $C$ follow from Equation \ref{eqn:eqn1} with an initial condition of $C = 0$. 

At $t=t_2$, $A$ switches off (A = 0 again) and similarly, the dynamics for $C$ follow from Equation \eqref{eqn:eqn1} with $C(t_2)$ as the initial condition. %as shown below:

Equation \ref{eqn:eqn1} can thus be  simplified to
\begin{align}
\label{eqn:eqn2a}
\begin{split}
\frac{dT_e}{dt} &= \beta_{T_e}\bigg( \frac{1}{K^n_{T_e} + 1} \times \frac{C^n}{K^n_{CT_e} + C^n}\bigg) - \alpha_{T_e} T_e,
\end{split}
\end{align}
where
\begin{equation}
\label{eqn:eqn3}
C = \begin{cases}
0, & \text{$t<t_1$},\\
\frac{\beta_c}{\alpha_c}\left(\frac{1}{K^n_{AC} + 1}\right)\left(1 - e ^{\alpha_c(t_1-t)}\right), & \text{$t_1\leq t< t_2$}.,\\
\frac{\beta_c}{\alpha_c}\left(\frac{1}{K^n_{AC} + 1}\right)\left(e ^{\alpha_c t_2} - e ^{\alpha_c t_1}\right)e ^{-\alpha_c t}, & \text{$ t>t_2$}.
\end{cases}
\end{equation}
Of note, coinhibition via competition for B7 molecules by CTLA-4 can be captured in $\alpha_{Te}$.

\subsection{The two-signal requirement with immune regulation  induce a coupled coherent-incoherent feedforward loop (CCI-FFL)}
\begin{figure}
\center
\includegraphics[scale=0.35]{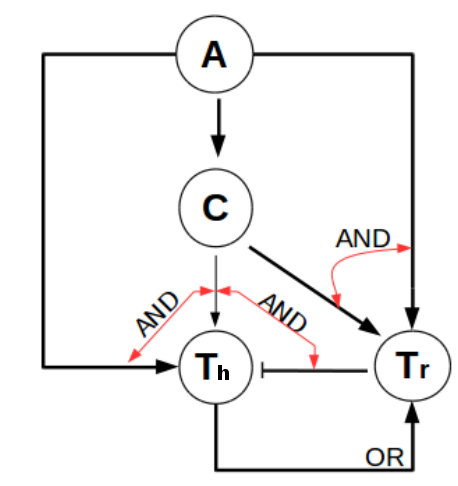}
\caption{Network structure of the coupled coherent-incoherent feed forward loop, proposed to govern helper-regulatory T cell  interactions. Initially, each T cell type is  activated by the two signals with the AND logic gate. Subsequent activations   are influenced by the statuses of the counter cell type.   }
\label{fig:b4}
\end{figure}

Usually, network motifs do not work in isolation: they are embedded within larger networks in a way
that preserves their dynamical functions \cite{alon2003biological}. The interaction between T helper ($T_h$) cells and T
regulatory ($T_r$) cells regulates the effector and regulatory properties of the immune system. An excessive
immune response by $T_h$ cells is normally prevented  by $T_r$ cells. Before it can suppress $T_h$ cells however, a $T_r$
cell must itself be activated following the same pathway like the T helper cell. In addition to this, $T_h$ cells
can enhance the activation and sustain proliferation of $T_r$ cells by producing a cytokine, interleukin 2 (IL-2)
\cite{savir2012balancing, guo2008cd28, refaeli1998biochemical}. IL-2, however, cannot substitute for CD28
in regulatory T cell activation \cite{hombach2007effective}. Therefore, as shown in Figure \ref{fig:b4}, the two-signal
requirement, combined with immune regulation, induce 
coupled coherent-incoherent feedforward Loop (CCI-FFL). Notably, this motif has an OR logic gate for the activation of $T_r$ cells.
This CCI-FFL is a multiple output feedforward loop since the antigenic and costimulation signals
regulate the activation of more than one type of T cell.
\subsubsection{Mathematical model of the coherent-incoherent feedforward loop}
Using the same approach as in Section \ref{sec:sec211}, we modeled the rates of change of $T_h$ and $T_r$ using the following system of differential equations:
\begin{align}
\label{eqn:eqn5}
\begin{split}
\frac{dT_h}{dt} &= \beta_{T_h}\bigg( \frac{A^n}{K^n_{AT_h} + A^n} \times \frac{C^n}{K^n_{CT_h} + C^n}\times \frac{K^n_{rh}}{K^n_{rh} + T_r^n}\bigg) - \alpha_{T_h} T_h,\\
\frac{dT_r}{dt} &= \beta_{T_r}\bigg( \frac{A^n}{K^n_{AT_r} + A^n} \times \frac{C^n}{K^n_{CT_r} + C^n}+ \theta \frac{T_h^n}{K^n_{hr} + T_h^n}\bigg) - \alpha_{T_r} T_r,
\end{split}
\end{align} 
where $C$ is  defined as in Equation \eqref{eqn:eqn3}. Equation \eqref{eqn:eqn5}
takes into account the OR input function for the activation of $T_r$ cells by $T_h$ produced IL-2; this activation is inadequate for maximal activation, hence the parameter $0<\theta<1$. The parameters $\beta$ represent  the maximal production rate of each entity, $K_{ij}$ are the activation/ suppression coefficient of entity $j$ by activating/ suppressing factor $i$ and $\alpha$ are the respective decay  rates.

Although it has been shown that regulatory T cells need not recognise
the same antigen as T helper cells \cite{alpan2004educated}, we considered the crossregulation model \cite{leon2000modelling} in which regulatory and helper T cells have same antigen specificity. All simulations were performed using the deSolve package in R and  the parameter values used are as shown in the respective figures in the results section.
\section{Results}
\subsection{The C1-FFL is a sign sensitive delay element in T cell activation.}
\begin{figure}
\center
\includegraphics[scale=0.9]{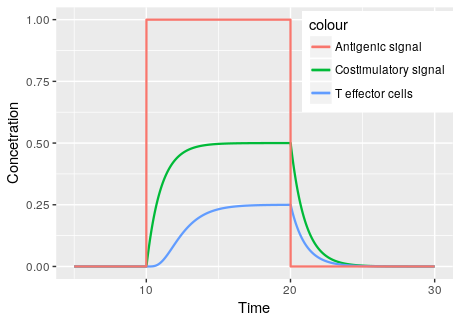}
\caption{Dynamics of C1-FFL with an AND gate input function as a sign sensitive delay element. While there is a lag to the activation of the T effector cell during the ON step, concentrations for both $C$ and $T_e$ begin to fall immediately when the antigenic signal is OFF. Parameters used were: $\beta_c = 1, \beta_e = 1, \alpha_c = 1, \alpha_e = 1, n = 4, K_{AC} = 1, K_{AT_e} = 1, K_{CT_e} = 0.5$. }
\label{fig:b2}
\end{figure}

Simulations based on the model in Equation \ref{eqn:eqn1} show that the coherent type 1 feedforward loop (C1-FFL) can serve as a sign-sensitive delay
element (Figure \ref{fig:b2}), that is, a circuit that responds with a delay to a step-like stimulus in one direction (e.g OFF to ON),
and rapidly to the step in the opposite direction (ON to OFF). The delay ($T_{ON}$ ) depends on the
activation threshold ($K_{CT_e}$ ) for the T cell as shown in Equation \ref{eqn:eqn4}.
This delay is the time  it takes $C$
to pass its activation threshold and is given by
\begin{eqnarray}
\label{eqn:eqn4}
T_{ON} = \frac{1}{\alpha_c}\log \bigg(\frac{C_{st}}{C_{st} - K_{CT_e}}\bigg),
\end{eqnarray}
where $C_{st}$ is the steady state concentration of the costimulation signal $C$. From this expression,
if $K_{CT_e} \rightarrow  C_{st}$, $T_{ON}\rightarrow \infty$, thus the activation threshold $K_{CT_e}$ should be  smaller than $C_{st}$ . Also, the larger the value of $K_{CT_e}$, the longer the delay (Supplementary Figure 2).
\subsection{The C1-FFL filters out noisy anitgenic signals.}
\begin{figure}
\center
\includegraphics[scale=0.9]{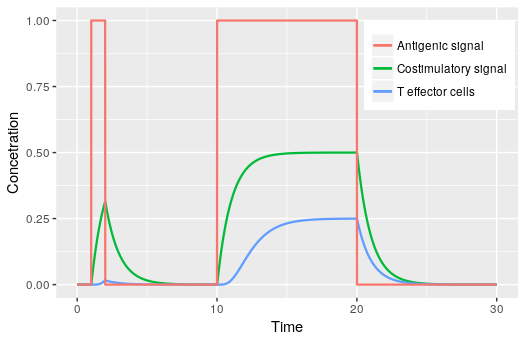}
\caption{Dynamics of C1-FFL with an AND gate input function as a noise detector. Short lived signals are unable to kick-start $T_e$ cell activation. Parameter values same as in Figure \ref{fig:b2}.}
\label{fig:b3}
\end{figure}

Model simulations also show that the C1-FFL motif filters out noisy signals (Figure \ref{fig:b3}). Long durations elicit a T cell response after a short delay whereas short durations fail to elicit a response. This failure  of T cell activation occurs because  the short duration  of the antigen signal does not give enough time for the costimulation signal to accumulate  past its
activation threshold. Thus, the C1-FFL motif filters  noise in the form of  short pulses of antigen signal  whilst  allowing persistent signal  to elicit T cell activation.

Of note,  while  we used a step-like stimulus in these simulations, sigmoidal forms  yield qualitatively similar dynamics. For example consideration of an exponentially growing antigen signal $A(t) = e ^{\lambda t}, \lambda > 0$, reproduces the delay in T cell activation as shown in Supplement Figure 3.%\ref{fig:b9}.

\subsection{The coupled coherent-incoherent feed forward loop is a sign sensitive delay element.}
\begin{figure}
\center
\includegraphics[scale=0.85]{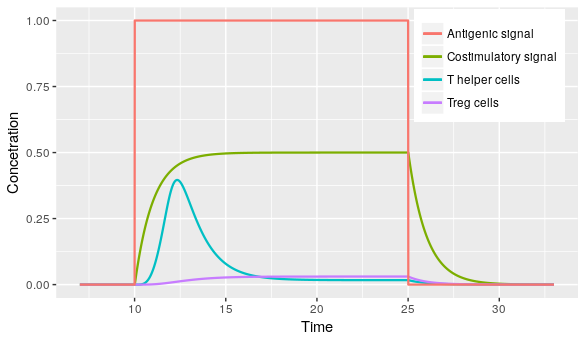}
\caption{Dynamics of coupled coherent-incoherent feedforward loop (CCI-FFL). Although activation of both T cell types delays, $T_r$ delays longer. An initial rise in the concentration of $T_h$ is suppressed by the rise of $T_r$. Parameters used were: $\beta_c = 1, \beta_r = 5, \beta_h = 50, \alpha_c = 1, \alpha_h = 1, \alpha_r = 1, n = 4, K_{AC} = 1,K_{CT_r} = 1.5, K_{AT_r} = 1,K_{AT_h} = 1, K_{CT_h} = 1, K_{hr}=10 , K_{rh} = 0.01$. }
\label{fig:b11}
\end{figure}

Simulations based on the mathematical model in Equation \ref{eqn:eqn5} show that the CCI-FFL retains the property of  sign sensitive
delay  (Figure \ref{fig:b11}). The delay in the activation of each T cell type depends mainly on their
respective costimulation threshold parameter, $K_{CT}$. In this simulation, the concentration of $T_h$ initially rises, unchecked, until $T_r$ passes its threshold and exerts a suppressive effect. We notice that when $T_h$ starts to drop, the concentration of $T_r$ is much lower than that of $T_h$. However, this also depends  on the simulation parameters. 
\begin{figure}
\center
\includegraphics[scale=0.85]{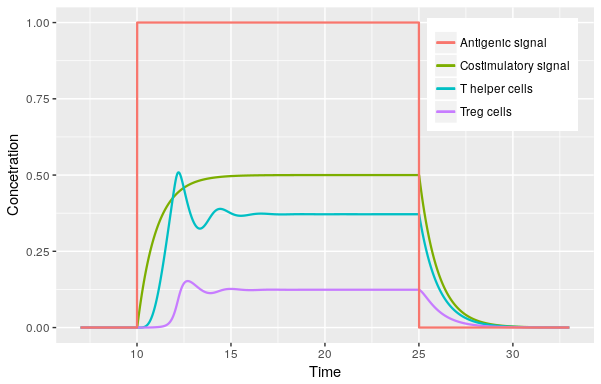}
\caption{Dynamics of coupled coherent-incoherent feedforward loop (CCI-FFL). Dampened oscillations in the concentration of activated effector and regulatory cells. Parameters used were: $\beta_c = 1, \beta_r = 5, \beta_h = 5, \alpha_c = 1, \alpha_h = 1, \alpha_r = 1, n = 4, K_{AC} = 1,K_{CT_r} = 1.5, K_{AT_r} = 1,K_{AT_h} = 1, K_{CT_h} = 0.5, K_{hr}=1 , K_{rh} = 0.1$. }
\label{fig:b10}
\end{figure}

\subsection{A condition for a healthy concentration of activated T helper cells  at equilibrium}
\label{sec:cond}
With n=1 for simplicity, analysis of the system in Equation \ref{eqn:eqn5}, at equilibrium, generates the following
condition for a healthy concentration of activated T helper cells:
\begin{eqnarray}
\label{eqn:eqn6}
\omega_h (K_{hr}+\varepsilon) <  \omega_r \left( \varepsilon + \frac{\varepsilon ^2}{K_{hr}}\right) +  K_{hr}K_{rh} \varepsilon + \left(K_{rh} + \theta\frac{ \beta_r}{\alpha_r}\right)\varepsilon ^2, 
\end{eqnarray}
where
\begin{eqnarray}
\omega_r &=& \frac{\beta_r}{\alpha_r}\left(\frac{A}{K_{AT_r}+A}\right) \left(\frac{C}{K_{CT_r}+C}\right) K_{hr},
\\\omega_h &=& \frac{\beta_h}{\alpha_h}\left(\frac{A}{K_{AT_h}+A}\right) \left(\frac{C}{K_{CT_h}+C}\right) K_{rh},
\end{eqnarray}
and $\varepsilon$ represents the maximum concentration of activated $T_h$ cells in normal conditions in the absence of any real  antigen. Parameters $\omega_r$ and $\omega_h$ represent the overall activation rates for $T_r$ and $T_h$ cells, respectively. Assuming $\varepsilon$ is very small, $\varepsilon ^2 \approx 0$, and Equation \eqref{eqn:eqn6} can be simplified to 
\begin{eqnarray}
\label{eqn:eqn7}
\omega_h (\frac{K_{hr}}{\varepsilon}+1) <  \omega_r  +  K_{hr}K_{rh}. 
\end{eqnarray}
Rearranging terms, we get:
\begin{eqnarray}
\label{eqn:eqn9}
\omega_h - \omega_r &<&   K_{hr}K_{rh} \left[1 - \frac{\beta_h}{\alpha_h \varepsilon}\left(\frac{A}{K_{AT_h}+A}\right) \left(\frac{C}{K_{CT_h}+C}\right)\right]\\
\omega_h - \omega_r &<& K_{hr}K_{rh}\phi
\end{eqnarray}
For a very small $\varepsilon$, and assuming $\beta_h\geq \alpha_h$, $\phi$ is likely to be negative considering typical values for $K_{AT_h}$ and $K_{CT_h}$.
In that case, Equation \eqref{eqn:eqn9} becomes
\begin{eqnarray}
\label{eqn:eqn8}
\omega_h + K_{hr}K_{rh}\phi <  \omega_r, 
\end{eqnarray}
which implies that 
\begin{eqnarray}
\label{eqn:eqn8}
\omega_h  <  \omega_r. 
\end{eqnarray}

\section{Discussion} 
 In this work, we showed that T cell activation can be modeled by a feed foward loop: a motif  capable of signal processing as a sign sensitive delay element. We argue that this is a desirable property  of any regulatory system of T cell activation as it ensures response to actual threats and minimises wastage after the threat has been eliminated.
 
The two-signal theory posits that T cells  require at least  two signals to become activated: one   in form of an antigenic signal (A) and the other  in the form of a costimulation signal (C). Investigating   the mode of T cell activation by these  signals shows that the two signal theory  induces a coherent type 1 feedforward loop. This network motif is known to behave as a  sign-sensitive delay element that responds  differently to signals of different signs. The
presence of a delay, $T_{ON}$, during the ON step enables  the T cell to decide whether or not to launch a response to
potential health threats. This delay depends inversely on the activation threshold, $K_{CT}$ as shown in Equation \ref{eqn:eqn4}. The higher the
threshold, the longer the delay. For very large values of $K_{CT}$, the T cell may never be activated given any duration of the antigen signal: This could be a plausible explanation for the phenomenon of self tolerance. On the other hand, the absence of a delay when the antigen signal is switched  off  ensures  less
wastage of resources  and  less damage to  tissues due to a prolonged response.

Additionally, the sign sensitive delay rejects any transient antigenic (A) signals and responds only to persistent signals.
This ensures T cell activation only when a signal is persistent (Figure \ref{fig:b3}). This is important because persistent signals are more likely to represent real
health threats compared with short lived signals, and therefore minimises the wastage of resources and the
side effects of a defective  response. T cell responses consume the body’s resources for instance ATP and so
if not required, these responses will be extremely inefficient. The two-signal model for T cell activation therefore
provides a molecular mechanism for ensuring that immune responses occur only when they are required
and for the correct amount of time. %This is useful because it allows the individual to conserve limited resources.

When the two-signal model was combined with the interaction between the $T_h$ and $T_r$ cells, we found that
this induces a coupled coherent-incoherent feedforward loop (CCI-FFL). The CCI-FFL forms a multiple output feedforward loop since antigenic signal (A) and costimulation
signal (C) regulate the activation of  more than one type of  cell. As the activation/inactivation signals vary through
time in the CCI-FFL, they pass through different activation/inactivation thresholds. Therefore, the CCI-FFL
can generate temporal orders of T cell activation/inactivation depending on the activation/inactivation
thresholds. %A similar temporal regulation of T helper and T regulatory  cells by antigen and costimulation exists.
 In Figure \ref{fig:b11}, we observe  delays in the initial activation of both $T_h$ and $T_r$ cells since they are individually governed by a C1-FFL . The difference in delay  is because $T_h$ and $T_r$ have different activation thresholds, and hence they wait for different lengths of
time after  the antigen signal has been turned on.
During infection,  the activation threshold for
$T_r$ is higher than that for $T_h $ \cite{hombach2007effective}, and so $T_h$ becomes activated first. The difference in delay should increase as
the difference $K_{CT_r} - K_{CT_h}$ increases (Equation \eqref{eqn:eqn4}). This makes sense biologically because we do not want $T_h$ to be
suppressed by $T_r$ before it has had the chance to clear an infection. Therefore, the difference between
$T_{ON} (T_h )$ and $T_{ON}(T_r )$ defines a critical time window within which $T_h $ can function unopposed by $T_r$. 

For a specific range of parameters, the dynamic model for the CCI-FFL can generate damped oscillations in the concentrations of $T_h$ and $T_r$ cells as shown in  Figure\eqref{fig:b10}. 
Inherent in the  CCI-FFL structure is a negative feedback loop with two nodes, i.e, $T_h$ and $T_r$. Negative feedback loops have been shown to be necessary for  periodic behavior \cite{snoussi1998necessary,gouze1998positive}. The oscillations   however, damp because repression by $T_r$ cells diminishes the activation and expansion of $T_h$ cells which, in turn, negatively affects the expansion of  $T_r$  via IL-2 activation. This effect is propagated through subsequent cycles as the two cell types approach equilibrium. This is useful because it allows the individual to conserve limited resource and regulate
the immune response corresponding to the infection level. For example, this can be helpful for infections
that have an on and off behaviour, and for chronic infections like HIV.

In the absence of an antigen, a small number  of  T helper cells, less than a threshold $\varepsilon$, are nevertheless activated. This could be, for example, in response to bacteria in the gut or any other commensal organisms in the body. In the presence of a real threat, we expect this number to increase in order to fight off the pathogen, and then to decrease once the threat has been eliminated. It is  the prolonged maintenance of such an increased quantity of activated $T_h$ cells that causes damage to body tissues and   is thus usually  suppressed by $T_r$ cells. A successful suppression of $T_h$ by $T_r$ will see this quantity fall back below $\varepsilon$. Using this bound on the concentration of activated $T_h$ cells under normal condition, we obtain a condition for a healthy concentration of activated T helper cells  at equilibrium, shown in Equation \eqref{eqn:eqn6}. A consideration of  $\varepsilon$, very small, as in  the case of self reacting $T_h$ cells, simplifies this condition to $\omega_h <  \omega_r.$ This shows that the overall activation rate of T regulatory cells should be greater than that for T helper cells for the former to successfully suppress $T_h$ cells and to prevent self reacting T helper cells. In particular the concentration of $T_r$ required for half-maximal repression of $T_h$ ($K_{rh}$) should ideally be much smaller than the concentration of $T_h$ required for half-maximal IL-2-mediated activation fo $T_r$ (i.e $K_{hr}$).

Understanding the organising principles of T cell activation is key to knowing how it works, when it will
be effective and why it sometimes fails, for example, why it is damped during cancer. In addition, it is also
important for devising new ways to therapeutically improve efficiency of T cell activation. 
In  this work, we applied network science to investigate the dynamical behaviour of T cell
regulation. We focused on the classical version of the two-signal theory, leaving the more complex
consideration of coinhibition for the future. We proposed  a novel structural representation of
the two signal theory of T cell activation in the form  of a coherent type 1 feedforward loop, and showed that it has desirable properties of an ideal immune response. Combining the
two signal requirement with immune regulation  induced a coherent-incoherent feedforward loop,   analysis of which  generated a condition  for  a healthy concentration of activated T helper cells  at equilibrium. 
  Our results provide new insight about how the two signal requirement ensures that T cell activation is
both targeted and efficient, and elucidate possible mechanisms for self tolerance.
\bibliographystyle{unsrt} \bibliography{references.bib}
\end{document}